\documentclass[twocolumn,aps,amsmath,amssymb,10pt]{revtex4-2}
\usepackage{graphicx}
\usepackage{dcolumn}
\usepackage{bm}

\usepackage{multirow}

\usepackage{amssymb}
\usepackage{amsmath}
\usepackage{graphicx}
\usepackage{xcolor}
\usepackage{braket}
\usepackage{CJK}
\usepackage{indentfirst}
\usepackage{amsmath}
\usepackage{cases}
\usepackage[T1]{fontenc}

\def\Eq#1{Eq.~(\ref{#1})}
\def\Fig#1{Fig.~\ref{#1}}

\newcommand{\<}{\langle}
\renewcommand{\>}{\rangle}

\newcommand{\bea}{\begin{eqnarray}}
\newcommand{\eea}{\end{eqnarray}}

\definecolor{UkiyoRed}{RGB}{223,126,102}
\definecolor{UkiyoGreen}{RGB}{148,181,148}
\definecolor{UkiyoYellow}{RGB}{237,199,117}
\definecolor{YinshuaiBlue}{RGB}{130,143,199}
\definecolor{YinshuaiPurple}{RGB}{134,128,207}
\definecolor{YinshuaiOrange}{RGB}{255,203,148}
\usepackage{hyperref}
\hypersetup{colorlinks=true, citecolor=blue, urlcolor=blue, linkcolor=blue}

\begin{document}
\title{Magnetic order and novel quantum criticality in the strongly interacting quasicrystals}

\author{Cong Zhang$^{1,2}$}
\author{Yin-Kai Yu$^{1,2}$}
\author{Shao-Hang Shi$^{1,2}$}
\author{Zi-Xiang Li$^{1,2}$}
\email{zixiangli@iphy.ac.cn}

\affiliation{$^1$Beijing National Laboratory for Condensed Matter Physics \& Institute of Physics, Chinese Academy of Sciences, Beijing 100190, China}
\affiliation{$^2$University of Chinese Academy of Sciences, Beijing 100049, China}

\date{\today}
\begin{abstract}
We present the sign-problem-free quantum Monte Carlo study of the half-filled Hubbard model on two-dimensional quasicrystals, revealing how specific aperiodic geometries fundamentally dictate quantum criticality.  By comparing the Penrose and Thue-Morse quasicrystals, we demonstrate that the nature of the magnetic phase transition is controlled by the electronic density of states (DOS): while the singular DOS of the Penrose tiling induces magnetic order at infinitesimal interaction strengths, the Thue-Morse lattice requires a finite critical interaction to drive the transition. 
Crucially, through a novel boundary construction strategy and rigorous finite-size scaling, we identify a quantum critical point on the Thue-Morse quasicrystal with critical exponents ($\nu \approx 0.94$, $\beta \approx 0.72$ and $z\approx 1.51$) that deviate significantly from the conventional $(2+1)$D Heisenberg $O(3)$ class. 
These findings establish the existence of a novel universality class driven by the interplay between electronic correlations and aperiodic geometry, challenging standard paradigms of magnetic criticality in two dimensions.
\end{abstract}
\maketitle

\pdfbookmark[1]{Introduction}{sec1}
\textcolor{YinshuaiBlue}{\it Introduction}--- Elucidating the quantum phases and phase transitions driven by strong electronic interactions constitutes a central theme in modern condensed matter physics. In particular, quantum criticality plays a pivotal role not only in statistical physics due to its universality~\cite{Sachdevbook,Sondhi1997RMP,Vojta2003Review,dqcp,li2017nc,Guo2016Science}, but also in various exotic phenomena in quantum many-body physics such as high-$T_c$ superconductivity and strange-metal transport~\cite{Abanov2003Review,Matsuda2014Review,Custers2003NatureQCP, RevModPhys.69.575}. 
While these phases and the associated criticality have been extensively explored in periodic systems, the corresponding physics in aperiodic environments—specifically quasicrystals—remains a largely unexplored frontier. 
Quasicrystals, characterized by long-range order without translational symmetry~\cite{PhysRevLett.53.1951}, host exotic single-particle states that are neither fully extended nor exponentially localized~\cite{albuquerqueTheoryElementaryExcitations2003, grimmUniversalLevelspacingStatistics2000,maciaRoleAperiodicOrder2006,PhysRevB.35.9529}. 
Instead, they typically exhibit critical, multifractal wavefunctions and singular densities of states (DOS)~\cite{godrecheMultifractalAnalysisReciprocal1990, gratiasSpatialOrderDiffraction2005, jagannathanFibonacciQuasicrystalCase2021a,PhysRevB.96.045138,Kohmoto1986PRBPenrose,Kohmoto1988PRBPenrose,PhysRevB.62.15569}. 
A fundamental open question is how these unique aperiodic electronic structures modulate quantum phase transitions and whether they support universality classes distinct from those established in conventional periodic systems.

Recent experimental breakthroughs have uncovered a rich landscape of many-body phenomena in quasicrystals~\cite{Kamiya2018NCSC, doi:10.1126/science.aar8412, PhysRevLett.120.140401, Yu2024, PhysRevX.15.011055, Kamiya2018,Terashima2024, PhysRevLett.122.110404, JeffrPNAS2020, PhysRevLett.125.200604,yaoprl2024}, including clear signatures of  magnetism and quantum criticality driven by strong correlations~\cite{Deguchi2012NMQCP,  TakePRL2023, Tokumoto2024, Tamura2025, PhysRevB.111.174409, LabibPRB2022,Rydh2025PRR}. Furthermore, the rapid development of moir\'e materials has revitalized this field, offering tunable platforms to explore electronic correlations in quasiperiodic potentials~\cite{Uri2023NatureMoire, doi:10.1073/pnas.1720865115, PhysRevB.99.165430,He2024NC, Lai2025}.  However, while various studies have addressed the many-body physics of quasicrystals~\cite{jagannathanQuasiperiodicHeisenbergAntiferromagnets2012,Fu2025arXivQC,PhysRevLett.123.210604,PhysRevLett.126.110401,PhysRevLett.134.136003,Gokmen2024, PhysRevLett.120.060407,Rydh2025PRRQCP, PhysRevLett.131.173402, PhysRevB.104.144511, PhysRevB.108.075121, PhysRevLett.109.106402, PhysRevLett.111.226401, PhysRevB.91.085125, PhysRevLett.116.257002, PhysRevX.6.011016,PhysRevLett.125.017002, Huang2020PRBAperiodic, chenHigherorderTopologicalInsulators2020, Yang2024PRL,PhysRevLett.121.126401,PhysRevB.88.125118,PhysRevB.95.024509, Yang2023NC, Huang2022PRL, PhysRevX.12.021058,Liu2025PRL,Huang2025arXiv,Yang2025arXiv2,   horiWeylSuperconductivityQuasiperiodic2024, PhysRevB.96.214402, chen2025quasicrystallinealtermagnetism, PhysRevB.92.224409}, rigorous investigations of symmetry-breaking orders and their associated critical properties using unbiased methods remain conspicuously scarce.  A systematic study of strongly correlated quasicrystals via non-perturbative approaches, specifically to unravel the nature of quantum phase transitions and the associated quantum critical points (QCPs), is therefore imperative. Such efforts will not only extend the fundamental theory of phase transitions beyond the paradigm of periodic lattices but also provide the essential theoretical framework needed to interpret the growing body of experimental discoveries

In this letter, we address this critical gap by presenting the first sign-problem-free projector quantum Monte Carlo (PQMC) study of the half-filled Hubbard model on two distinct two-dimensional aperiodic tilings~\cite{Sorella1989EPL,AssaadReview,li2019review,bercxALFAlgorithmsLattice2017}: the Penrose quasicrystal and the Thue-Morse quasicrystal. Despite the Hubbard model serving as the paradigmatic framework for strongly correlated systems~\cite{Qin2022Review,Berg2022Review,Huang2017Science,Huang2017Science,Zhang2024Science,Qin2020PRX,He2025PRL,Xu2025SciPost,Mei2025arXiv}, its ground-state phase diagram on quasicrystals has remained unexplored via unbiased numerical approaches. Our results reveal a striking dichotomy dictated by the underlying geometry. 
On the Penrose tiling, the presence of macroscopically degenerate confined states leads to a divergent DOS at the Fermi level, inducing long-range N\'eel order at infinitesimal interaction strengths. In contrast, the Thue-Morse quasicrystal possesses a finite DOS and undergoes a quantum phase transition from a disordered metal to a N\'eel ordered state only at a finite critical interaction $U_c$. 
Most notably, our large-scale QMC simulations and finite-size scaling analysis of the Thue-Morse transition yield critical exponents  that deviate significantly from the conventional $(2+1)$D $O(3)$ universality class. These findings provide robust evidence for a novel class of quantum criticality driven by the interplay of strong correlations and aperiodic geometry.

\begin{figure}[t]
	\centerline{\includegraphics[scale=0.28]{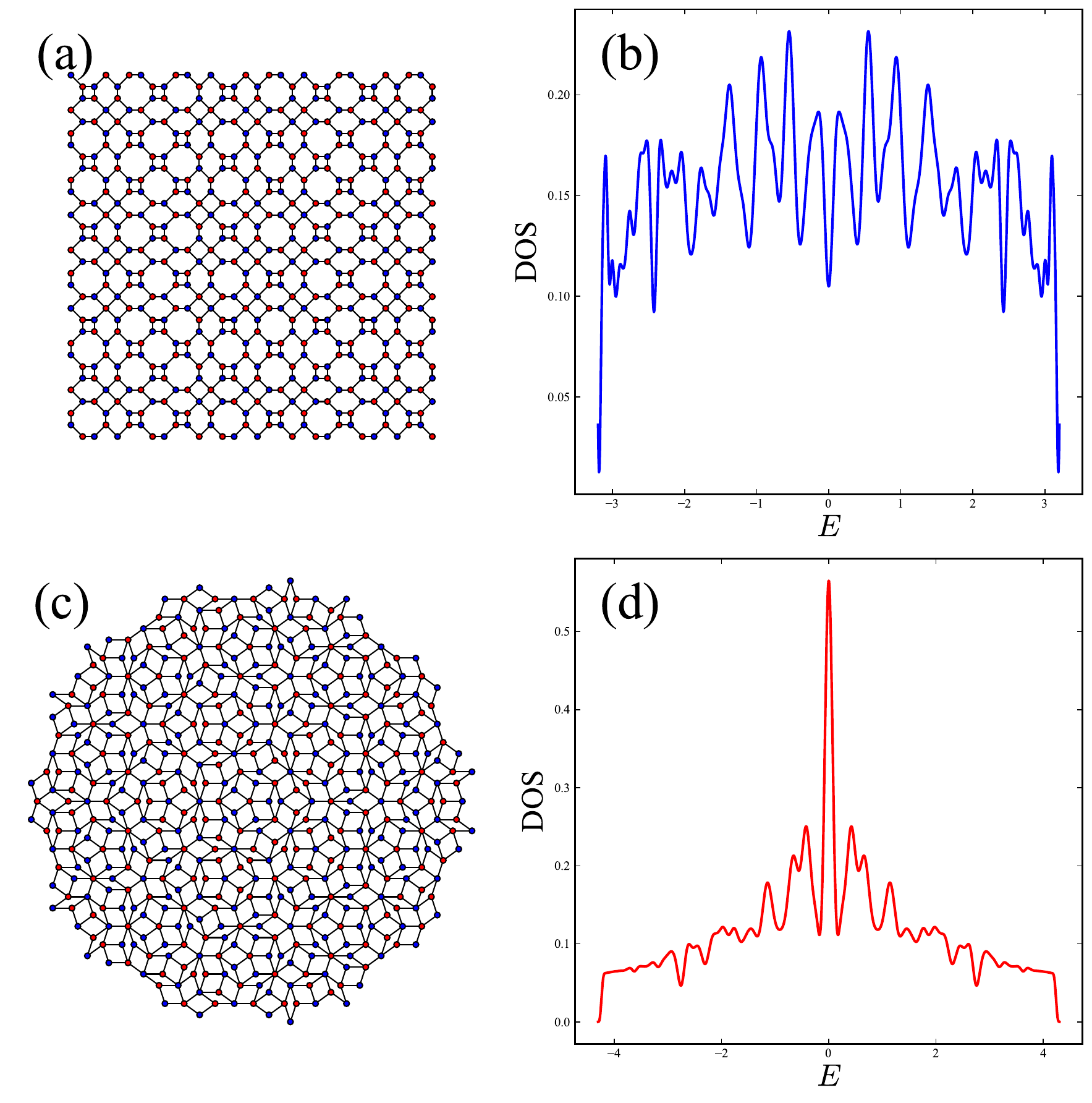}}
	\caption{(a) Schematic representation of the two-dimensional Thue-Morse quasicrystal. 
(b) DOS for the Thue-Morse tight-binding model. 
Calculations were performed using the KPM on a supercell of $8,388,608$ sites. The DOS remains finite at the Fermi level ($E=0$). 
(c) Schematic representation of the two-dimensional Penrose quasicrystal. 
(d) DOS for the Penrose tight-binding model for a cluster of $N_s = 6,738,718$ sites. 
In contrast to the Thue-Morse case, a divergent peak is observed at the Fermi level.}
	\label{Fig1}
\end{figure}

\pdfbookmark[1]{Model and Method}{sec2}
\textcolor{YinshuaiBlue}{\it Model and Method}---
We consider the half-filled Hubbard model, governed by the Hamiltonian:
\bea
\hat{H}  =& -t \sum_{\langle i, j\rangle, \sigma}\left(\hat{c}_{i \sigma}^{\dagger} \hat{c}_{j \sigma} + \text { H.c. }\right) \nonumber\\
&+U \sum_{i} (\hat{n}_{i \uparrow} - 1/2)( \hat{n}_{i \downarrow}-1/2),\label{eq1}
\eea
where $\hat{c}_{i \sigma}$ denotes the electron annihilation operator for spin $\sigma$ at lattice site i, $\hat{n}_{i \sigma}=\hat{c}^{\dagger}_{i \sigma}\hat{c}_{i \sigma}$ is the particle number operator. In addition, $t$ denotes the nearest-neighbor hopping amplitude (set to unity unless otherwise specified) and $U>0$ represents the on-site Coulomb repulsion. The Hubbard model serves as a prototypical framework for describing repulsive electronic interactions. 
In the strong-coupling limit, the half-filled model maps onto the antiferromagnetic Heisenberg model, rendering it an ideal platform for investigating the emergence of magnetic order and quantum phase transitions. Here, we study this model on two specific two-dimensional geometries: the Penrose tiling and the Thue-Morse quasicrystal~\cite{1974The,debruijnAlgebraicTheoryPenroses1981a,WOLNY2000313,PhysRevB.37.4375,morettiTwodimensionalPhotonicAperiodic2007}. The Penrose tiling represents a canonical two-dimensional quasicrystal with fivefold ($D_5$) rotational symmetry, while the Thue-Morse quasicrystal is a two-dimensional generalization of the one-dimensional Thue-Morse sequence, possessing $D_2$ or $D_4$ symmetry. As illustrated in Figs.~\ref{Fig1}(a) and \ref{Fig1}(c), both the rhombus Penrose tiling and the Thue-Morse lattice are bipartite. Furthermore, we employ the Kernel Polynomial Method (KPM)~\cite{RevModPhys.78.275} to calculate the DOS of ultra-large systems approaching the thermodynamic limit, with details provided in the Supplementary Materials (SM). As shown in Figs.~\ref{Fig1}(b) and (d), our calculations reveal a finite DOS at the Fermi level for the Thue-Morse quasicrystal, in sharp contrast to the divergent DOS observed in the Penrose tiling.

To accurately determine the ground-state properties of this strongly correlated model, we employ the unbiased PQMC method. 
Crucially, because both the Penrose tiling and the Thue-Morse quasicrystal are bipartite lattices, the Hubbard models defined in \Eq{eq1} are free from the fermion sign problem~\cite{Loh1990PRB,Wu2005PRBsign,Li2015PRBsign,Li2016PRLsign,xiang2016prl,Han2024arXiv,Wei2024PRBsignproblem,Chang2024PRB,Mondaini2022science,wang2015prl,Berg2012science,Meng2022PRB,Mondaini2025arXivsign,Yu2024arXiv}. This absence allows for efficient, unbiased PQMC simulations on large system sizes, yielding high-precision data essential for analyzing quantum phase transitions and critical phenomena. Further details regarding the PQMC algorithm and the conditions for avoiding the sign problem are provided in the SM. To investigate the magnetic order induced by the Hubbard interaction and the nature of the quantum phase transition, we evaluate the squared staggered magnetization, $M_2$, and the associated Binder ratio, $U_L$. These quantities are defined as:
\bea
M_2 &=& \frac{1}{N_s^2} \sum_{i, j} \epsilon_{i} \epsilon_{j} \<\hat{S}_{i} \cdot  \hat{S}_{j} \>\\
M_4 &=& \frac{1}{N_s^4} \sum_{i, j, k, l} \epsilon_{i} \epsilon_{j} \epsilon_{k} \epsilon_{l} \<(\hat{S}_{i} \cdot  \hat{S}_{j}) (\hat{S}_{k} \cdot  \hat{S}_{l})\> \\
U_{L}&=& \frac{M_2^2}{M_4}.
\label{eq2}
\eea
Here, $\epsilon_i = \pm 1$ denotes the parity of the sublattice to which site $i$ belongs, and $N_s$ is the total number of lattice sites. Throughout this article, $L$ is defined as $\sqrt{N_s}$. In the thermodynamic limit, $M_2$ converges to the square of the N\'eel order parameter. The Binder ratio $U_L$ serves as a powerful indicator for characterizing long-range order and the associated phase transition~\cite{Binder1981}. Theoretically, $U_L \to 1$ in the N\'eel ordered phase and $U_L \to 0.6$ in the disordered phase in the thermodynamic limit. At the putative quantum critical point (QCP), $U_L$ becomes scale-invariant (independent of system size) as it is a dimensionless quantity.

\begin{figure}[t] 
    \centering
    \includegraphics[width=\linewidth]{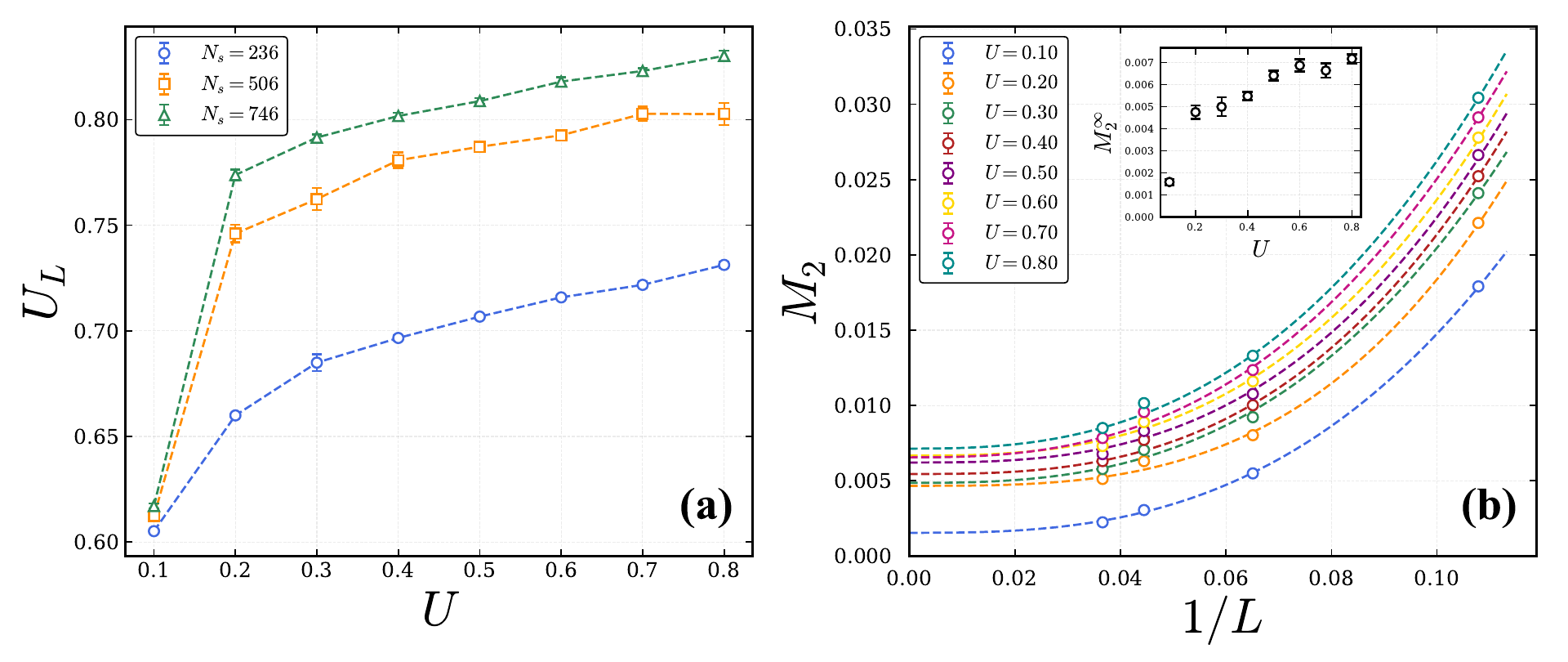}  
    \caption{
The Neel order on Penrose Quasicrystals. (a) Variation of the Binder ratio $U_L$ with $U$. The colors of the curves represent different system sizes $L$. In the range $U > 0$, $U_L$ systematically increases with increasing $L$ and gradually saturates. (b) Extrapolation of the squared order parameter $M_2$ with $1/L$. Different symbols correspond to different $U$ values, and the dashed line denotes the power function fit, where $f=ax^b+c$. Inset: Variation of $M_2$ in the thermodynamic limit with $U/t$, indicating the immediate emergence of non-zero magnetization for $U > 0$.
    }
    \label{Fig2}
\end{figure}

\pdfbookmark[1]{Penrose quasicrystal}{sec3}
\textcolor{YinshuaiBlue}{\it Penrose quasicrystal}---
To elucidate the fundamental influence of the quasicrystalline electronic structure on correlation effects, we first investigate the Penrose quasicrystal. Fig.~\ref{Fig2} displays the squared staggered magnetization $M_2$ and the Binder ratio $U_L$ as a function of the interaction strength $U$ for various system sizes. We fix the particle number at half-filling; in this regime, the DOS diverges at the Fermi level due to the existence of macroscopically degenerate confined states~\cite{Kohmoto1986PRBPenrose,Kohmoto1988PRBPenrose}. The Binder ratio results, depicted in \Fig{Fig2}(a), show that $U_L$ increases with system size across the entire regime of finite $U$, with no crossing points observed. This strongly indicates that an infinitesimal Hubbard interaction drives the system into a N\'eel ordered state. Moreover, the finite-size scaling of $M_2$ to $L \to \infty$, shown in \Fig{Fig2}(b), confirms a finite N\'eel order parameter in the thermodynamic limit for all considered values of $U$. Therefore, we conclude that for the Penrose lattice at half-filling, the ground state possesses long-range N\'eel order throughout the entire interaction regime. This behavior stems from the infinite DOS at the Fermi energy, which leads to a divergence of the magnetic susceptibility toward N\'eel ordering.

\pdfbookmark[1]{Thue-Morse quasicrystal}{sec4}
\textcolor{YinshuaiBlue}{\it  Thue-Morse quasicrystal}---
In contrast to the Penrose quasicrystal, the half-filled Thue-Morse quasicrystal possesses a finite DOS at the Fermi energy. This property renders it a potential platform for realizing novel quantum criticality separating the N\'eel ordered state from the disordered state as the interaction strength $U$ increases. To accurately quantify the antiferromagnetic quantum phase transition on the Thue-Morse quasicrystal, we comprehensively analyze the finite-size scaling of the dimensionless Binder ratio $U_L$ and the squared staggered magnetization $M_2$. However, unlike periodic systems, aperiodic lattices typically suffer from severe finite-size effects. We address this challenge by developing a boundary construction strategy analogous to the rational periodic approximation~\cite{entin-wohlmanPenroseTilingApproximants1988}. As detailed in the SM, this method involves selecting specific truncation positions to ensure the boundary environment adheres to the Thue-Morse generation rules. This approach significantly suppresses artifacts arising from boundary defects, thereby allowing for a precise finite-size scaling analysis.

\begin{figure}[t]
    \centering
	\includegraphics[width=\linewidth]{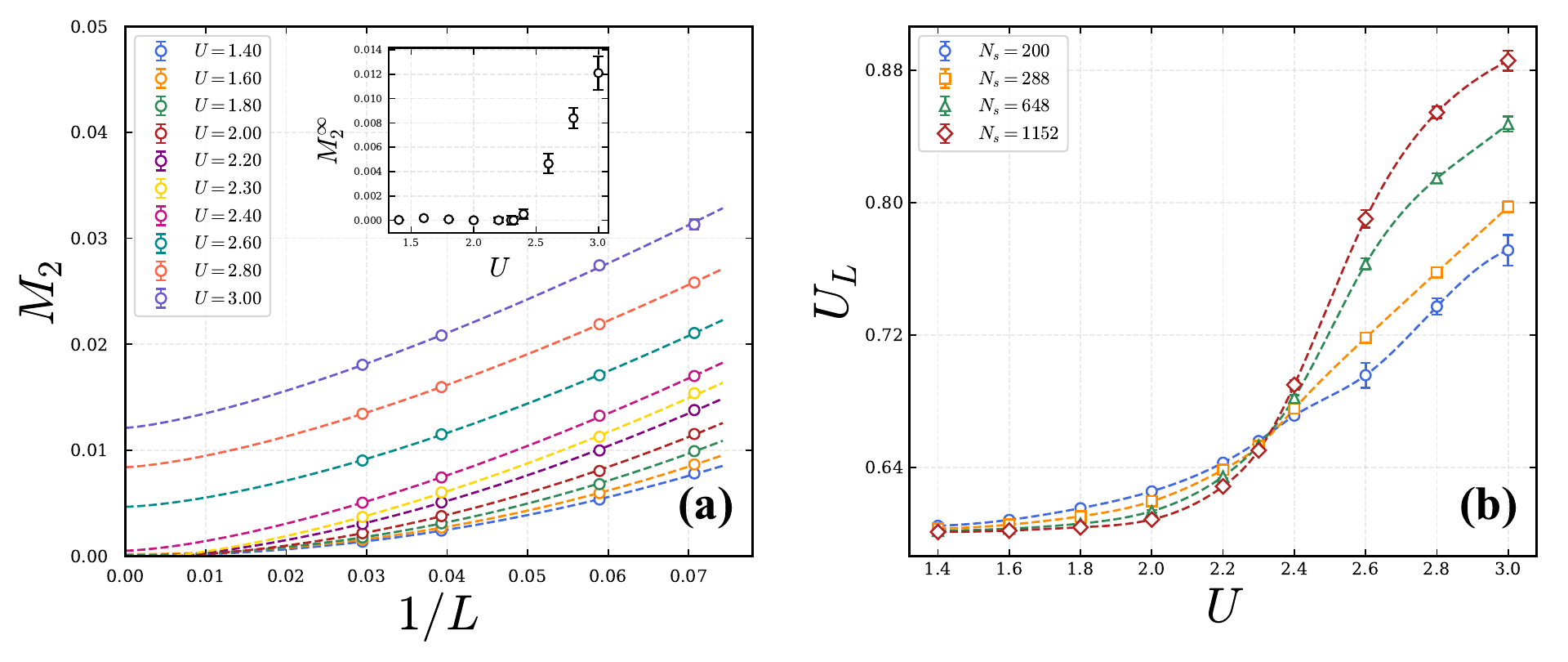}
	\caption{(a) Finite-size scaling of the squared staggered magnetization $M_2$ plotted against the inverse system size $1/L$ for various interaction strengths $U$. 
The dash lines represent  power function fit to the thermodynamic limit ($1/L \to 0$). 
Inset: The extrapolated $M_2$ in the thermodynamic limit, $M_{2}^{\infty}$, as a function of $U$. 
It shows a continuous onset of magnetic order at the critical point $U_c$. 
(b) The Binder ratio $U_L$ as a function of $U/t$ for different system sizes $L$. 
The crossing point of these curves identifies the critical interaction strength $U_c$.}
	\label{Fig3}
\end{figure}

The main panel of Fig.~\ref{Fig3}(a) displays $M_2$ as a function of the inverse system size $1/L$ for various interaction strengths $U$. By performing  power function fit on the data  to the thermodynamic limit ($1/L \to 0$), we extract the squared order parameter, $M_{2}^{\infty}$. As shown in the inset of Fig.~\ref{Fig3}(a), the dependence of $M_{2}^{\infty}$ on $U$ exhibits the signature of a continuous phase transition. Below the critical value $U_c/t \approx 2.32$, the order parameter vanishes, corresponding to a disordered phase. Above this threshold, the order parameter increases continuously, signaling the onset of long-range antiferromagnetic order.

Significantly, the Binder ratio data presented in Fig.~\ref{Fig3}(b) provides unambiguous evidence of a continuous transition from a disordered phase to a magnetically ordered N\'eel state. For $U < U_c \approx 2.32t$, $U_L$ decreases with increasing system size, indicating the absence of long-range order. Conversely, for $U > U_c$, $U_L$ increases with system size, signaling the emergence of long-range antiferromagnetism. At the critical point $U = U_c$, the $U_L$ curves for different system sizes intersect at a single point, pinpointing the location of the QCP. This critical value, $U_c \approx 2.32 t$, is consistent with the result obtained from the extrapolation of $M_2$, confirming the robustness of our analysis. Consequently, our results establish the ground-state phase diagram of the Hubbard model on the Thue-Morse quasicrystal, demonstrating that the Hubbard interaction drives a quantum phase transition at $U \approx 2.32 t$.

Furthermore, recent symmetry analyses suggest that N\'eel order on the Thue-Morse lattice, which possesses $D_4$ point group symmetry, corresponds to altermagnetic order~\cite{Yang2025arXiv2}, where the two sublattices are related by lattice rotational symmetries rather than translation or inversion alone~\cite{AM}. Consequently, our QMC results provide unambiguous evidence for the emergence of long-range altermagnetic order and its associated QCP in the Hubbard model on the Thue-Morse lattice. This constitutes a robust, unbiased numerical demonstration of altermagnetism within a microscopic model of a quasicrystal. A systematic investigation into the detailed properties, including its specific electronic and magnetic structures, remains a promising avenue for our future studies.

\begin{figure*}[htbp]
    \centering   
    \includegraphics[width=\textwidth]{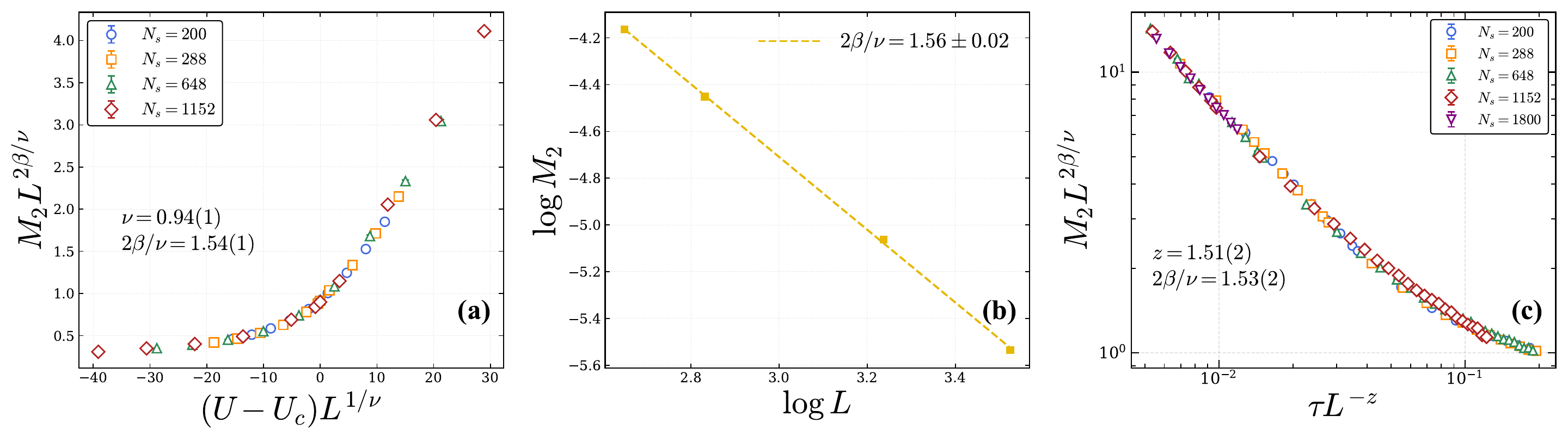}
    \caption{
(a) Finite-size scaling data collapse of the squared staggered magnetization $M_{2}$. 
The rescaled $M_{2} L^{\frac{2\beta}{\nu}}$ is plotted against the scaling variable $(U-U_{c})L^{1/\nu}$. 
Using the critical exponents $\nu=0.94(1)$ and $\frac{2\beta}{\nu}=1.54(1)$, the data for different system sizes collapse onto a single universal curve.  (b) Log-log plot of $M_{2}$ versus system size $L$ at the critical point $U_c = 2.32(1)$. 
The dash line represents a linear fit, yielding a slope of $\frac{2\beta}{\nu}=1.56(2)$, which is consistent with the result from the collapse analysis in (a). (c) Non-equilibrium finite-size scaling data collapse of $M_2$. Using the critical exponents $z=1.51(2)$ and $\frac{2\beta}{\nu}=1.53(2)$, the data for different system sizes collapse onto a single universal curve.
    }
    \label{Fig4}
\end{figure*}

\pdfbookmark[1]{Novel quantum criticality on Thue-Morse quasicrystal}{sec5}
\textcolor{YinshuaiBlue}{\it Novel quantum criticality on Thue-Morse quasicrystal}--- Next, we perform a systematic finite-size scaling analysis to elucidate the quantum critical properties of the magnetic transition on the Thue-Morse quasicrystal. 
In the vicinity of a putative QCP, the squared order parameter $M_2$ obeys the scaling form $M_2(U,L) = L^{-\frac{2\beta}{\nu}} \mathcal{F}[(U-U_c)^{\frac{1}{\nu}} L]$, where $\nu$ is the correlation-length exponent and $\frac{\beta}{\nu}$ corresponds to the scaling dimension of the magnetic order parameter. 
Consequently, with the appropriate choice of critical exponents $\nu$ and $\beta$, the rescaled data $M_2 L^{\frac{2\beta}{\nu}}$ plotted as a function of the scaling variable should collapse onto a single universal curve. 
We perform this data collapse analysis and obtain the critical exponents $\nu = 0.94(1)$ and $\frac{2\beta}{\nu}=1.54(1)$, as shown in \Fig{Fig4}(a).
Furthermore, the high quality of the data collapse provides robust evidence for a QCP separating the disordered and N\'eel ordered states. We further verify the critical properties by performing a direct scaling analysis of the squared staggered magnetization $M_2$ at the QCP.  At fixed $U=U_c$, the power-law dependence of $M_2$ on the system size $L$ manifests as a linear relationship in a log-log plot. 
A linear fit to this slope yields an exponent ratio of $\frac{2\beta}{\nu} = 1.56(2)$, as depicted in \Fig{Fig4}(b). 
This value is consistent with the result obtained from the data collapse analysis  within the error bars, confirming the robustness of the critical exponents derived from our finite-size scaling.

Furthermore, we determine the dynamic exponent $z$, which characterizes the dispersion of the order-parameter excitations, using a newly developed short-time relaxation approach within the PQMC framework~\cite{yu2023dirac,Yu2024arXiv}. Unlike standard equilibrium PQMC, this method efficiently extracts critical exponents, particularly $z$, during the non-equilibrium stage of imaginary-time evolution (details provided in the SM). We analyze the evolution of $M_2$ and $U_L$ initialized from a fully polarized N\'{e}el state. At the critical point $U=U_c$, $M_2$ follows the scaling form $M_2(\tau,L) = L^{-\frac{2\beta}{\nu}}\mathcal{G}(\tau L^{-z})$, where $\tau$ denotes the imaginary time~\cite{Yins2014prb}. A data collapse analysis of $M_2$ versus 
$\tau L^{-z}$ yields $z=1.51(2)$ and $\frac{2\beta}{\nu} =1.53(2)$, as shown in \Fig{Fig4}(c). The result for $\frac{2\beta}{\nu}$ is consistent with the equilibrium ground-state 
scaling results in \Fig{Fig4}(a), validating the reliability of our analysis. 
Similarly, scaling analysis of the Binder ratio $U_L$ yields a consistent value for 
$z$ (see SM). Additionally, a direct finite-size analysis of $M_2$ versus $\tau$ 
during short-time relaxation yields consistent values of $\frac{2\beta}{\nu z} = 1.01(3) $, 
further corroborating the extracted exponents. Consequently, the determined critical 
exponents [$\nu = 0.94(1)$, $\beta = 0.72(1)$ and $z=1.51(2)$] deviate significantly from those of the conventional $(2+1)$D Heisenberg 
$O(3)$ universality class ($\nu \approx 0.71$, $\beta \approx 0.37$, and 
$z=1$)~\cite{Campostrini2002prb}.

In periodic systems, the coupling between the magnetic order parameter and gapless fermionic excitations typically fundamentally alters the critical properties of the transition. 
On the Thue-Morse quasicrystal, gapless excitations also exist in the disordered metallic phase prior to the onset of N\'eel order. 
According to the standard Hertz-Millis-Moriya theory for two-dimensional metals, the dynamical critical exponent $z=2$ leads to an effective dimension $d+z=4$, which is the upper critical dimension of $\phi^4$ theory; consequently, the transition should be characterized by mean-field exponents~\cite{Hertz1976prb,Millis1993prb}. 
However, our observed critical exponents are clearly inconsistent with mean-field predictions. 
Another potential scenario is the Gross-Neveu-Yukawa universality class (chiral Heisenberg class), where criticality is modified by gapless Dirac fermions~\cite{Rosenstein1993PLB,Herbut2006PRL,Assaad2013prx,Li2015njp,Scherer2017PRD,Li2018SA,Lang2019prl,Wessel2016prb,Abolhassan2022PRL,Li2024PRLNon-Hermitian,Wang2014njp}. 
However, the physics here differs in two key respects: first, unlike Dirac fermions which exhibit a vanishing DOS at the node, the Thue-Morse fermions possess a finite DOS. Second, our obtained exponents [$\nu = 0.94(1)$, $\beta = 0.72(1)$ and $z=1.51(2)$] differ significantly from those of the chiral Heisenberg transition of SU(2) Dirac fermions on periodic lattices [$\nu = 1.02(1)$, $\beta = 0.76(2)$ and $z=1$]~\cite{Sorella2016prx}. 
Consequently, the QCP discovered in our study lies beyond the known universality classes of periodic fermionic systems. 

While the presence of gapless fermions with a finite DOS naturally distinguishes the Thue-Morse transition from the conventional bosonic $(2+1)$D Heisenberg $O(3)$ class, the deviation from standard fermionic QCP on periodic systems suggests a deeper mechanism rooted in aperiodicity. 
Unlike Bloch waves in periodic lattices, the low-energy electronic states in quasicrystals possess unique multifractal characteristics~\cite{jagannathanFibonacciQuasicrystalCase2021a}. 
Specifically, the intensity distribution of these critical wavefunctions exhibits spatial singularities, which fundamentally reorganizes local quantum fluctuations. 
This non-uniform spatial structure likely modifies the renormalization group flow, thereby altering the divergence of the correlation length and the resulting critical exponents. 
Thus, the novel QCP identified here is an emergent phenomenon originating from the interplay between strong correlations and the unique geometry of quasicrystalline wavefunctions. 
A rigorous quantitative field-theoretical derivation of these exponents remains an exciting challenge for future study.

\pdfbookmark[1]{Concluding remarks}{sec6}
\textcolor{YinshuaiBlue}{\it Concluding remarks}--- In this letter, we present the first sign-problem-free QMC study to systematically investigate quantum phase transitions in strongly correlated electronic models on quasicrystals. 
Our unbiased simulations reveal that on the Penrose quasicrystal, an infinitesimal Hubbard interaction is sufficient to drive N\'eel order, which then persists throughout the entire interaction regime. 
In contrast, a finite critical interaction is required to trigger the N\'eel transition on the Thue-Morse quasicrystal. This fundamental difference originates from the DOS at the Fermi level: while the Penrose tiling exhibits a divergent DOS due to macroscopic confined states, the Thue-Morse DOS remains finite. The N\'eel magnetic phase revealed in the Thue-Morse lattice is identified as an altermagnetic phase according to the lattice symmetry analysis. Most significantly, our large-scale simulations identify a QCP on the Thue-Morse lattice with critical properties that are fundamentally distinct from the conventional $(2+1)$D Heisenberg $O(3)$ universality class. The novel magnetic QCP emerging on the Thue-Morse lattice is distinct from the known universality classes of the quantum phase transition on periodic systems, arising from the interplay between strong electronic interactions and unique electronic structures in quasicrystal. 

Our discovery of a novel QCP in quasicrystals transcends the current understanding of criticality derived from periodic systems, opening a new frontier in the study of quantum phase transitions. At present, a general theoretical framework for quantum criticality in aperiodic environments remains conspicuously absent. Consequently, developing systematic analytical field theories or applying numerical renormalization group methods to quasiperiodic systems represents a crucial and promising direction for future research to pursue.

{\em Note added}: Upon completing this work, we became aware of an interesting related work by Ji, Shao, Liu and Yang that also
studies the Hubbard model on bipartite quasicrystal by projector quantum Monte-Carlo~\cite{Yang2025arXivQC}. The primary focus of their studies differs from ours.

\textcolor{YinshuaiBlue}{\it Acknowledgments}---  We sincerely thank Fan Yang for helpful discussions. This work is supported by the National Natural Science Foundation of China under Grant Nos. 12347107 and 12474146, and Beijing Natural Science Foundation under Grant No. JR25007.

%

\onecolumngrid
\newpage
\widetext
\thispagestyle{empty}

\setcounter{equation}{0}
\setcounter{figure}{0}
\setcounter{table}{0}
\renewcommand{\theequation}{S\arabic{equation}}
\renewcommand{\thefigure}{S\arabic{figure}}
\renewcommand{\thetable}{S\arabic{table}}

\pdfbookmark[0]{Supplementary Materials}{SM}
\begin{center}
    \vspace{3em}
    {\Large\textbf{Supplementary Materials for}}\\
    \vspace{1em}
    {\large\textbf{Magnetic order and quantum criticality in Hubbard model on quasicrystal}}\\
    \vspace{0.5em}
\end{center}

\section{Kernel Polynomial Method}
In this work, the DOS of quasicrystal tight-binding model is calculated using the Kernel Polynomial Method based on Chebyshev polynomial expansion. The core idea is to expand the spectral function of the target operator (e.g., the Hamiltonian $\hat{H}$) in the basis of Chebyshev polynomials $T_n(x)$. For a Hamiltonian normalized to the interval $\left[-1, 1\right]$, the DOS can be approximated as:
\begin{equation}
    \rho(E) \approx \frac{1}{\pi \sqrt{1-E^{2}}}\left[\mu_{0}+2 \sum_{n=1}^{N_{c}} \mu_{n} T_{n}(E)\right],
\end{equation}
where the expansion coefficient $\mu_n=\operatorname{Tr}\left[T_n(\hat{H})\right]$ contains the spectral information of the system. To efficiently estimate the trace of large-dimensional matrices, we adopt the random vector technique, i.e.,
\begin{equation}
    \operatorname{Tr}\left[T_{n}(\hat{H})\right] \approx \frac{1}{R} \sum_{r=1}^{R}\langle r| T_{n}(\hat{H})|r\rangle,
\end{equation}
where $|r\rangle$ is a random vector. To suppress Gibbs oscillations introduced by high-order expansions, the above expansion is multiplied by a convergence kernel (the Jackson kernel is used in this work). In this study, we set the truncation order of the Chebyshev expansion to $N_c = 100$ and use $R = 100$ random vectors for trace estimation, thereby achieving a good balance between computational convergence and accuracy.

\section{The Details of Two-Dimensional Thue-Morse Quasicrystals}\label{sec:A1}
This Section details the construction method of the two-dimensional Thue-Morse Quasicrystal employed in this work, with a focus on illustrating the periodic boundary conditions adapted to the intrinsic structure of the quasicrystal.

\subsection{The construction of Two-Dimensional Thue-Morse Quasicrystal}
The Thue-Morse quasicrystal is constructed from the Thue-Morse sequence. Given the ($N-1$)-th sequence $S_{N-1}$, the $N$-th sequence is formed by concatenating with its complementary sequence $[S_{N-1}]$, namely $S_{N} = S_{N-1}[S_{N-1}]$
, and it typically starts with $S_{0}=P$, $[S_{0}]=M$
. This rule can be naturally extended to two dimensions: by defining $P$ and $M$ as two basic unit cells with different geometric configurations and arranging them spatially according to the aforementioned sequence rules, a two-dimensional Thue-Morse lattice structure can be constructed. In a square lattice, this can be expressed as
\begin{equation}
    S_{N}=\left|\begin{array}{cc}
{\left[S_{N-1}\right]} & S_{N-1} \\
S_{N-1} & {\left[S_{N-1}\right]}
\end{array}\right|.
\end{equation}
This structure inherits the intrinsic self-similarity and long-range quasi-order of the sequence but completely lacks translational symmetry.

\subsection{The Realization of Periodic Boundary Conditions}
For such aperiodic systems, conventional periodic boundary conditions (PBCs)—which directly connect opposite edges—forcibly introduce crystalline translational symmetry, disrupting the intrinsic correlations of quasicrystals and leading to severe boundary effects that invalidate finite-size scaling analysis.

To enable reliable finite-size simulations of infinite aperiodic structures, we have developed a periodization construction method based on the intrinsic generation rules of the sequence. The core of this method lies in the following: the simulated finite system must be a self-consistent segment of the original Thue-Morse (TM) sequence, and its boundary connections must strictly restore the inherent adjacency relations in the sequence—thereby preserving the integrity of the quasicrystal’s local structure even under periodic repetition.

We illustrate this construction with a specific sequence example. Consider the 3rd-generation Thue-Morse sequence: 
\begin{equation}
S_3=PMMPMPPM
\end{equation}
We extract its first 5 characters PMMPM as the basic periodic unit in both the x and y directions. The rationality of imposing PBCs must be justified by verifying that the connection between the two ends of the segment is consistent with the rules in the infinite sequence.

For instance, the infinite sequence along the x direction is presented as follows :
\begin{equation}
    \dots PMMPM|{\color{red}P}PM\dots
\end{equation}
the character immediately following the right-end character M of the segment is exactly P; meanwhile, the starting character at the left end of the segment is precisely P. Thus, connecting the right-end M to the left-end P at the periodic boundary accurately reproduces the generation rule P followed by M in the TM sequence. Conversely, for the left-end character P, we only need to consider a longer preceding subsequence like
\begin{equation}
    S_4 = PMMPMPPMMPPMPMMP
\end{equation}
When extending the sequence to the left, the left neighbor of the left-end character of the selected segment is found to be M, as illustrated by the extended sequence :
\begin{equation}
    S_5 = MPPMPMMPPMMPMPP{\color{red}M}|PMMPM|{\color{red}P}PMMPPMPMMP
\end{equation}
which matches the right-end M of the segment. This logic can be recursively applied to verify the connections of all boundary points via the sequence’s generation rules. The situation along the y direction is consistent with the above.

Therefore, the PBCs defined in this manner ensure that the nearest-neighbor environment of every lattice site—especially boundary sites—in the finite system is identical to that of the corresponding site in the infinite TM quasicrystal. This method fundamentally mitigates unphysical finite-size effects introduced by artificial truncation, providing a crucial technical guarantee for extracting accurate critical scaling behaviors from finite-size data.

\section{Projector Quantum Monte Carlo algorithm}\label{sec:A2}
To accurately solve the ground-state properties of the repulsive Hubbard model on quasicrystal, we employed the Projector Quantum Monte Carlo (PQMC) method. The core idea is to start from a trial wavefunction $\left|\psi_{T}\right\rangle$
with non-zero overlap with the ground state, and project the true ground state $\left|\psi_{0}\right\rangle$
by applying a sufficiently long imaginary-time evolution operator 
$e^{-\tau \hat{H} }$:
\begin{equation}
    \left|\psi_{0}\right\rangle \propto \lim _{\tau \rightarrow \infty} e^{-\tau \hat{H}}\left|\psi_{T}\right\rangle .
\end{equation}
The ground-state expectation value of an observable $\langle \hat{O} \rangle$ can be calculated via the following formula:
\begin{equation}
    \langle \hat{O} \rangle = \frac{\langle \psi_T | e^{-\tau \hat{H}} \hat{O} e^{-\tau \hat{H}} | \psi_T \rangle}{\langle \psi_T | e^{-2\tau \hat{H}} | \psi_T \rangle} .
    \label{S7}
\end{equation} 

The total projection time $2\tau$ is discretized into $L_{\tau}$ time slices, i.e., $\Delta_{\tau}=2 \tau / L_{\tau}$. At each time slice, the kinetic and potential terms are first separated using the second-order symmetric Trotter–Suzuki decomposition:
\begin{equation}
    e^{-\Delta_{\tau}\hat{H}} \simeq e^{-\frac{\Delta_\tau}{2} \hat{H}_{t}} e^{-\Delta_{\tau} \hat{H}_{U}} e^{-\frac{\Delta_{\tau}}{2} \hat{H}_{t}} + \mathcal{O}(\Delta_\tau^3) ,
\end{equation}
with
$\hat{H}_{t} = -t \sum_{\langle i, j\rangle, \sigma} \left( \hat{c}_{i \sigma}^{\dagger} \hat{c}_{j \sigma} + \text { h.c. } \right)$ being the quadratic hopping term and $
\hat{H}_{U} =U \sum_{i} (\hat{n}_{i \uparrow} - 1/2)( \hat{n}_{i \downarrow}-1/2)$ being the quartic Hubbard interaction term.

Subsequently, we apply the Hubbard-Stratonovich (HS) transformation 
to decouple the on-site Hubbard interaction , introducing 
an Ising-type auxiliary field $s_l(i)$ at each site $i$ and time slice $l$:
\begin{equation}
    \mathrm{e}^{-\Delta_{\tau} U \left(\hat{n}_{i \uparrow}-1 / 2\right)\left(\hat{n}_{i \downarrow}-1 / 2\right)}= C \sum_{s_{l}(i) = \pm 1} \mathrm{e}^{ i \alpha s_{l}(i) \left(\hat{n}_{i \uparrow}+\hat{n}_{i \downarrow} - 1\right)},
\end{equation}
where $C=\mathrm{exp}{(\Delta_{\tau} U / 4)}/2$ and $\cos{\alpha} = \mathrm{e}^{-\Delta_{\tau} U / 2}$. 
The HS decomposition is performed in the density channel and the full spin $SU(2)$ symmetry of the original Hubbard model is explicitly preserved. After the transformation, the fermionic degrees of freedom become quadratic and can be integrated analytically. As a result, the dominator in in Eq.~(\ref{S7}) can be expressed as a sum over the weights of all possible configurations of the auxiliary field:
\begin{equation}
    \langle \psi_T | e^{-2\tau \hat{H}} | \psi_T \rangle = C^m \sum_{\{ s_l(i) = \pm 1\}} \langle \psi_T | \prod_{l=1}^{L_{\tau}} \mathrm{e}^{-\Delta_{\tau}\hat{H}_t/2} \mathrm{e}^{i\alpha \sum_{i} s_l(i) (\hat{n}_{i\uparrow}+\hat{n}_{i\downarrow}-1)} \mathrm{e}^{-\Delta_{\tau}\hat{H}_t/2} | \psi_T \rangle  = \sum_{\mathbf{s}} W_{\mathbf{s}}
\end{equation}
where $\mathbf{s} = \{s_l(i) = ±1\}$ denotes the set of auxiliary fields. 
Finally, the expectation value $\langle \hat{O} \rangle$ can be rewritten as a weighted average over auxiliary–field configurations:
\begin{equation}
    \frac{\langle \psi_T | e^{-\tau \hat{H}} \hat{O} e^{-\tau \hat{H}} | \psi_T \rangle}{\langle \psi_T | e^{-2\tau \hat{H}} | \psi_T \rangle}  =  \frac{ \sum_{\mathbf{s}}   W_{\mathbf{s}}  \langle O \rangle_{\mathbf{s}}}{\sum_{\mathbf{s}} W_{\mathbf{s}}}
\end{equation}
We can rigorously prove that, at half filling on a bipartite quasicrystalline
lattice, the auxiliary-field quantum Monte Carlo simulations are free of the
sign problem. Under a partial particle--hole transformation $\hat{c}_{i\downarrow} \rightarrow (-1)^{i} \hat{c}^{\dagger}_{i\downarrow}$, the kinetic part $\hat{H}_{t}$ remains invariant. If, in addition, the chosen trial wave function, represented by a Slater determinant $|P\rangle$, is such that the spin-up $|P^{\uparrow}\rangle$ and spin-down sectors $|P^{\downarrow}\rangle$ become identical
after the transformation, consequently, the weight $W_{\mathbf{s}}$ can be factorized into a product of two complex-conjugate parts:
\begin{equation}
\begin{aligned}
W_{\mathbf{s}} & \propto \langle \psi_T^{\uparrow} |  \prod_{l=1}^{L_{\tau}} \mathrm{e}^{-\Delta_{\tau}\hat{H}_t^{\uparrow}/2} \mathrm{e}^{i\alpha \sum_{i} s_l(i) \hat{n}_{i\uparrow}} \mathrm{e}^{-\Delta_{\tau}\hat{H}_t^{\uparrow}/2}  |\psi_T^{\uparrow} \rangle     \langle \psi_T^{\downarrow} |  \prod_{l=1}^{L_{\tau}} \mathrm{e}^{-\Delta_{\tau}\hat{H}_t^{\downarrow}/2} \mathrm{e}^{-i\alpha \sum_{i} s_l(i) \hat{n}_{i\downarrow}} \mathrm{e}^{-\Delta_{\tau}\hat{H}_t^{\downarrow}/2}  |\psi_T^{\downarrow}  \rangle\\
& \propto W_{\mathbf{s}}^{\uparrow} W_{\mathbf{s}}^{\downarrow} = |W_{\mathbf{s}}^{\uparrow}|^2 \geq 0.
\end{aligned}
\end{equation}
The problem is thus mapped onto the sampling of an effective single-particle
problem in the auxiliary-field configuration space.
We perform an ergodic Markov-chain Monte Carlo sampling of the auxiliary fields
using a local Metropolis update scheme.
After equilibration, physical observables are measured over a series of
statistically independent auxiliary-field configurations. Detailed derivations can be found in this review \cite{AssaadReview}.

The Hubbard model in this study is at half-filling and defined on a bipartite lattice. To ensure the simulation is free of the sign problem, we adopted specific trial wavefunction construction strategies tailored to the electronic structures of different quasicrystals.
For the Thue-Morse quasicrystal, we directly adopt the ground-state Slater determinant wavefunction at half-filling as the trial wavefunction $\left|\psi_{T}\right\rangle$. This accelerates the convergence to the limit of $\tau \rightarrow \infty $ with
much less computational effort.
For the Penrose quasicrystal, its non-interacting tight-binding model is strictly gapless at the Fermi level, with a divergent density of states.  To lift the degeneracy, we introduce a small staggered magnetic
field in the non-interacting Hamiltonian $\hat{H}_t$,
\begin{equation}
    \hat{H}_t \rightarrow \hat{H}_t + h \sum_i (-1)^i \hat{S}_i^z ,
    \label{S15}
\end{equation}
with $\hat{S}^{z}_{i} = \hat{c}^{\dagger}_{i\uparrow}\hat{c}_{i\uparrow} - \hat{c}^{\dagger}_{i\downarrow} \hat{c}_{i\downarrow}$ and $h \ll 1$.
The ground state of the modified Hamiltonian is then taken as the trial wave function. We emphasize that this modification of the trial wave function is introduced solely for numerical stabilization. Moreover, we have confirmed that the $h$ implemented in our simulation is small enough to have a negligible impact on the numerical results. 

To ensure projection to the true ground state, we set the total projection time $\tau =50$ and verified that key physical quantities have fully converged under this parameter. To obtain reliable results, our data are averaged over 10–40 independent Markov chains. After thermalization, each Markov chain undergoes 50–240 measurement sweeps. Our time slices $\Delta_{\tau}$ are set to values around 0.05, which ensures that the results obtained are sufficiently  stable. All calculations are implemented based on the open-source software package ALF(Algorithms for Lattice Fermions)~\cite{bercxALFAlgorithmsLattice2017}, which provides a rigorously validated PQMC algorithm framework, ensuring the reliability and reproducibility of our simulations.

In addition, in order to extract the dynamical critical exponent $z$ at the
quantum critical point, we employ the developed short imaginary-time relaxation dynamical approach\cite{yu2023dirac,Yu2024arXiv}.
In this framework, $\tau$ is reinterpreted as the imaginary-time relaxation
duration. We choose a fully polarized N\'eel-ordered state as the initial state for the imaginary-time evolution, which corresponds to taking the staggered field in Eq.~\ref{S15} to the limit $h \to \infty$.
Accordingly, this choice also preserves the absence of the sign problem.

\section{Short-time non-equilibrium imaginary-time critical dynamics}

\begin{figure*}[bt]
    \centering
    \includegraphics[width=0.95\textwidth]{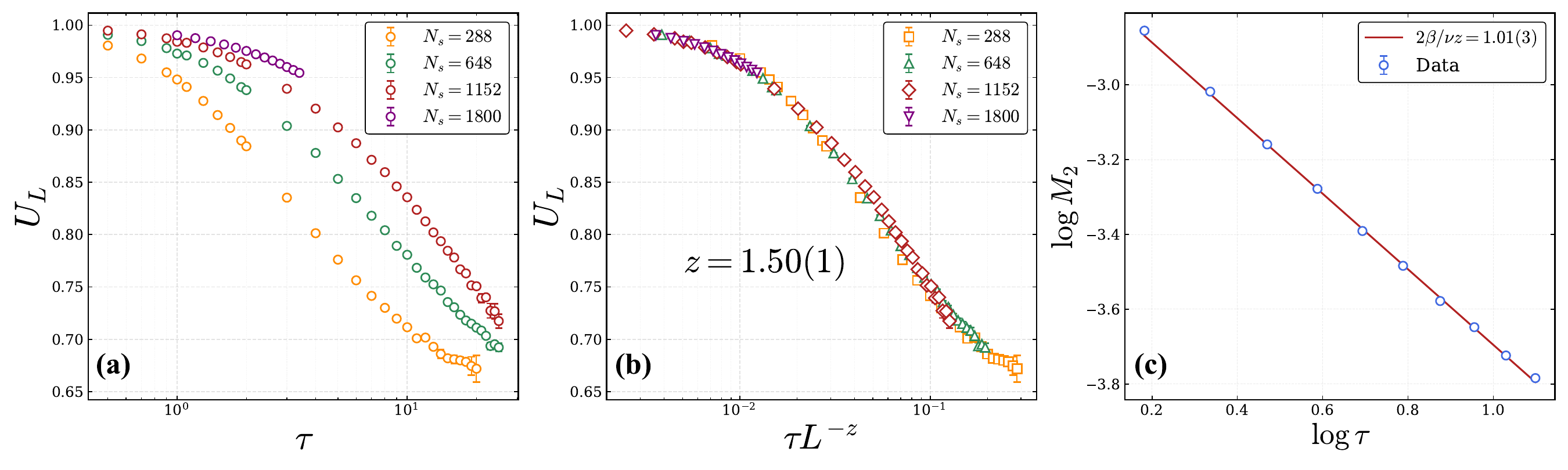}
    \caption{Imaginary-time critical relaxation dynamics starting from the fully polarized N\'{e}el state on the Thue-Morse quasicrystal with $U=U_c$. (a) $U_L$ as a function of the projection imaginary time $\tau$ for different numbers of lattice sites $N_s$. (b) Data collapse for panel (a), with $z = 1.50(1)$. (c) Power-law decay of $M_2$ with $N_s=1800$ in the short-time stage, $M_2 \propto \tau^{-\frac{2\beta}{\nu z}}$, yielding $\frac{2\beta}{\nu z} = 1.01(3)$.}
    \label{fig:SM-ITCD}
\end{figure*}

To investigate the novel quantum criticality emerging in the Thue-Morse quasicrystal, 
we extract the critical exponents, specifically the dynamic exponent $z$, from the 
imaginary-time non-equilibrium relaxation dynamics. When the projection time $\tau$ 
is not sufficiently large, the system has not yet converged to the ground state and 
remains in a non-equilibrium regime. As systematically demonstrated in previous works, 
this non-equilibrium imaginary-time evolution provides an efficient and reliable 
theoretical framework for determining the critical properties of quantum many-body 
systems via PQMC. According to the non-equilibrium scaling theory of quantum 
criticality~\cite{Yins2014prb,yu2023dirac}, if the initial state corresponds to a 
fixed-point wave function, the finite-size scaling form of the Binder ratio $U_L$ 
generalizes from the conventional equilibrium form 
$U_L(U,L) = \mathcal{G}\left((U-U_c)L^{\frac{1}{\nu}}\right)$ to
\begin{equation}
    U_L(\tau,U,L) = \mathcal{G}_1\left(\tau L^{-z}, (U-U_c)L^{\frac{1}{\nu}}\right).
\end{equation}
We fix the interaction strength to the critical value $U = U_c$, and the scaling form then reduces to
\begin{equation}
    U_L(\tau,L) = \mathcal{G}_2(\tau L^{-z}).
    \label{eq:ITCD-Binder}
\end{equation}
We choose the trial wavefunction $\left|\psi_{T}\right\rangle$ as a fully polarized N\'{e}el state, tracking how the system evolves as the projection imaginary time $\tau$ increases.  Numerically, the critical relaxation processes of the Binder ratio $U_L(\tau)$ for different system sizes are shown in Fig.~\ref{fig:SM-ITCD}(a). According to Eq.~\eqref{eq:ITCD-Binder}, we rescale the imaginary time $\tau$ by $L^{z}$, and obtain the data collapse shown in Fig.~\ref{fig:SM-ITCD}(b), which yields $z = 1.50(1)$, consistent with the value $z = 1.51(2)$ obtained in the main text.

Similarly, for the non-equilibrium state, the finite-size scaling form of the squared order parameter $M_2$ is also generalized from the conventional form
$M_2(U,L) = L^{-\frac{2\beta}{\nu}}\mathcal{F}\left((U-U_c)L^{\frac{1}{\nu}}\right)$ to
\begin{equation}
    M_2(\tau,U,L) = L^{-\frac{2\beta}{\nu}}\mathcal{F}_1\left(\tau L^{-z}, (U-U_c)L^{\frac{1}{\nu}}\right).
\end{equation}
At the critical point $U = U_c$, one has
\begin{equation}
    M_2(\tau,L) = L^{-\frac{2\beta}{\nu}} \mathcal{F}_2(\tau L^{-z}).
    \label{eq:ITCD-M2-1}
\end{equation}
In the main text, we rescale $M_2(\tau)$ for different system sizes according to Eq.~\eqref{eq:ITCD-M2-1}, and obtain $\frac{2\beta}{\nu} = 1.53(2)$ and $z = 1.51(2)$ from the optimal data collapse shown in Fig.~\ref{Fig4}(c).
Furthermore, by redefining $\mathcal{F}_3(\tau L^{-z}) \equiv (\tau L^{-z})^{\frac{2\beta}{\nu z}}\mathcal{F}_2(\tau L^{-z})$, the relaxation scaling form at the critical point can be equivalently written as
\begin{equation}
    M_2(\tau,L) = \tau^{-\frac{2\beta}{\nu z}} \mathcal{F}_3(\tau L^{-z}).
    \label{eq:ITCD-M2-2}
\end{equation}
In the short-time stage, namely for small projection imaginary time $\tau$, the function $\mathcal{F}_3(\tau L^{-z})$ in Eq.~\eqref{eq:ITCD-M2-2} only needs to be expanded to zeroth order, yielding $M_2(\tau) \propto \tau^{-\frac{2\beta}{\nu z}}$.
We identify this power-law decay stage in the large system with $N_s = 1800$ (see Fig.~\ref{fig:SM-ITCD}(c)), and perform a linear fit on the log-log scale, obtaining $\frac{2\beta}{\nu z} = 1.01(3)$, which is also consistent with our previous results.

\end{document}